\documentclass[ reprint, amsmath,amssymb,aps, pre, groupedaddress]{revtex4-2}

\usepackage{bm, amsmath, amsfonts, mathtools, physics}

\usepackage{ifpdf}
\ifpdf
  \usepackage{graphicx}
\else
  \usepackage[dvipdfmx]{graphicx}
\fi

\usepackage{ulem}
\usepackage{here}
\usepackage{comment}

\usepackage{color}
\newcommand{\REV}[1]{\textcolor{black}{#1}}

\newcommand\erase{\bgroup\markoverwith{\textcolor{red}{\rule[.5ex]{2pt}{0.8pt}}}\ULon}

\begin{document}

\title{Anti-aligning Self-propelled Model of Two Species: Emergence of \textit{Self-organized Heterogeneous Aligned and Clustered Order} 
}

\author{Takahiro Oki$^{1}$}
\email{Contact author: toki-doki-t3@g.ecc.u-tokyo.ac.jp}
\author{Tetsuhiro S. Hatakeyama$^{2}$}
\author{Seiya Nishikawa$^{3}$} 
\author{Shuji Ishihara$^{3,4}$}
\author{Toshinori Namba$^{3,4,}$}
\email{Contact author: nmb@g.ecc.u-tokyo.ac.jp}

\affiliation{$^{1}$Department of Physics, The University of Tokyo, 7-3-1 Hongo, Bunkyo-ku, Tokyo 113-0033, Japan}
\affiliation{$^{2}$Earth-Life Science Institute (ELSI), Institute of Science Tokyo, 2-12-1-IE-1 Ookayama, Meguro-ku, Tokyo 152-8550, Japan}
\affiliation{$^{3}$Graduate School of Arts and Sciences, The University of Tokyo, Komaba 3-8-1, Meguro-ku, Tokyo 153-8902, Japan}
\affiliation{$^{4}$Universal Biology Institute, The University of Tokyo, Komaba 3-8-1, Meguro-ku, Tokyo 153-8902, Japan}

\begin{abstract}  
Self-propelled particles with anti-aligning interactions generally do not form a polar order. 
However, in this Letter, we show that when multiple types of such particles coexist and interact through aligning interactions between different species, a global polar order can emerge through the formation of elongated clusters with alternating domains of each species. 
By developing a mean-field theory, we reveal the conditions for cluster formation and characterize the resulting patterns.
Our findings highlight the critical role of inter-species interactions in the emergence of complex ordered~states.
\end{abstract}
\maketitle


Collective motion is widely observed in various biological systems, such as birds~\cite{Ballerini2008}, fish~\cite{Berdahl2013}, insects~\cite{Buhl2006}, bacteria~\cite{Zhang2010}, and cytoskeletal filaments~\cite{Sanchez2012, Sumino2012}, and manifests self-organization in living systems.
Such systems show a global polar order that spontaneously emerges through local interactions without external control.
It is currently a widely studied topic in non-equilibrium statistical physics~\cite{Ramswamy2010, Marchetti2013, Gompper2020, Toner2024}. The Vicsek model is the most pioneering and representative model~\cite{Vicsek1995}, in which self-propelled particles align with their neighbors, leading to the formation of a polar order and collective movement in a cluster.
Besides the Vicsek model, various other models have been proposed to study collective motion, most of which incorporate aligning interactions~\cite{Chaté2020}.

In contrast, even though the anti-aligning interaction has been recognized as a source of rich physical phenomena in systems other than those of active particles, e.g., the spin and neural network systems~\cite{Binder1986,Mezard2009}, the behavior of active particles with
anti-aligning interactions has not been sufficiently studied.
One possible reason is the traditional belief that such interactions lead to a disordered state, that is, the particles move randomly and do not form a coherent structure. 
Indeed, recent studies have demonstrated that the anti-aligning Vicsek model (AAVM) does not show a global polar order across the entire system, although repetitive formation and destruction of hexagonal patterns of small particle clusters~\cite{Escaff2024,Escaff2024b} are observed.
The extension of the AAVM to a two-species model with anti-aligning interactions between the species results in a global polar order within each species, but not across species~\cite{Kursten2023}.
These results challenge the traditional belief that anti-aligning interactions cannot lead to organized structures and raise the following question: How can self-propelled particles achieve a global polar order across an entire system despite the presence of anti-aligning interactions?

In this Letter, we introduce multiple species of anti-aligning self-propelled particles and demonstrate that a global polar order emerges through aligning interactions between different types of particles.
In the steady state, the different particles form an elongated cluster with a stripe pattern. 
Within the cluster, the particles are aligned in the same direction along the long axis, causing the cluster to move ballistically along that axis.
We term this polar and clustered state as the \textit{Self-organized Heterogeneous Aligned and Clustered Order} (SHACO). 
For the emergence of the SHACO state, the aligning interaction between the different types of particles does not need to be larger than the anti-aligning interaction between the same type of particles. 
The stability of the SHACO state can be explained using the mean-field analysis, which can also effectively describe the characteristics of the stripe pattern and transition between the disordered and SHACO states.
In previous studies of pattern formation in multi-species systems, non-reciprocal interactions between different particles were mainly investigated~\cite{Grauer2020,Fruchart2021,Ouazan-Reboul2023,Dinelli2023,Mandal2024}, but we found that even if the interaction is reciprocal, multi-species of self-propelled particles will show a characteristic pattern.

{\it Model.---} 
We consider self-propelled particles comprising two distinct populations, named type A and type B, in two dimensions. 
The particles either tend to align or anti-align with each other depending on their type. 
The number of self-propelled particles of type A and B are $N_A$ and $N_B$, respectively.
The position of the $i$-th particle at time $t$ is represented by the vector $\bm{r}_i(t) = (x_i(t),y_i(t))$, and the particle moves in the direction $\theta_i(t)$ with a constant velocity subject to noise. 
The dynamics of each particle are the same as those in the time-continuous Vicsek model.
\begin{align}
\dot{\bm{r}_i} &= v_0\bm{s}_i + \sqrt{\sigma}\bm{\xi}_i\ , \label{eq:dot_ri}\\
\tau \dot{\theta}_i &= - \langle m_{ij}\sin{(\theta_i - \theta_j)} \rangle_{R}\ . \label{eq:dot_theta_i}
\end{align}
Here ${\bm s}_i \equiv (\cos \theta_i, \sin \theta_i)$ is a unit vector representing the direction of self-propulsion, 
$\bm{\xi}_i$ represents a Gaussian white noise of zero mean and unit variance, and $\sigma$ is the dispersion of the noise.
$v_0$ is the self-propelled velocity and is normalized to unity by selecting an appropriate time unit.
The angle brackets $\langle \cdots \rangle_{R}$ in Eq.~\eqref{eq:dot_theta_i} denote the average over the neighboring particles $j$ that are within the distance of $R$ from the particle $i$.
The angle $\theta_i$ tends to align or anti-align with the neighboring particles $j$ depending on the sign of $m_{ij}$: 
The $i$-th particle aligns (anti-aligns) with the $j$-th particle for a positive (negative) $m_{ij}$.
$m_{ij}$ is set to $m_{AA}$ ($m_{BB}$) if both particles are of type A (type B) and to $m_{AB}$ or $m_{BA}$ if the particles in the interacting pair are of different types.
$m_{AB}$ is generally distinct from $m_{BA}$ ($m_{AB} \neq m_{BA}$), but in this study we restricted our analysis to the symmetric case of $m_{AA} = m_{BB}$ and $m_{AB} = m_{BA}$, i.e., the system is reciprocal. 
$\tau$ is the ratio of the characteristic timescales of directional rotation to particle motion.
Because we consider the anti-aligning interaction between the same type of particles, we can set $m_{AA} = m_{BB} =-1$ without loss of generality, and set $m_{AB}=m_{BA}=c$, where $c$ is in the range of $0 \leq c \leq 1$.
We term this model as the two-species AAVM. 
For comparison, we also conducted a numerical simulation of the one-species AAVM~\cite{Escaff2024}; 
see the supplementary material (SM) Sec.~S1~\cite{supp}.

We numerically simulated the above Eqs.~\eqref{eq:dot_ri} and \eqref{eq:dot_theta_i} in a two-dimensional $L \times L$ square region under periodic boundary conditions. 
The parameters were set as $v_0=1.0, \tau =0.05$, $R = 3.0$, $N_A=1,000$, $N_B = 1,000$, $L=32$, and the noise strength $\sigma = 0.04$, unless otherwise noted.
The initial conditions were randomly chosen such that both the position ${\bm r}_i$ and angle $\theta_i$ were uniformly distributed.

\begin{figure}[tbhp]
\includegraphics[width=0.95\linewidth]{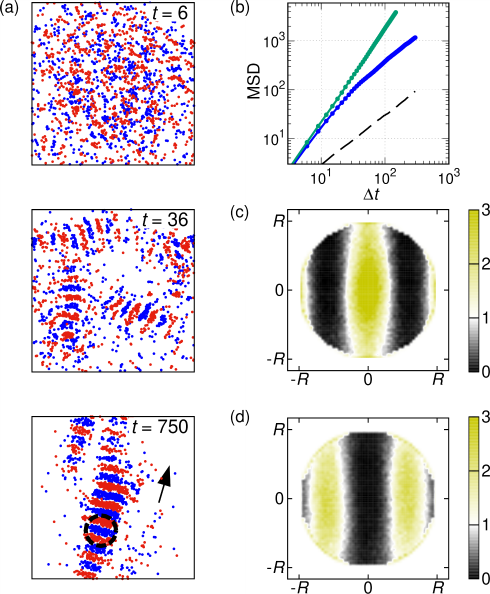}
\caption{
Behavior of the two-species AAVM at $c=1.0$. (a) Snapshots at early, middle, and late stages (from top to bottom).
Type A and B particles are represented by red and blue dots, respectively.
The black dashed circle indicates the interaction region of radius $R$. 
The black arrow indicates the forward direction of the cluster.
(b) The mean squared displacement (MSD) of a particle in the steady state.
Green, blue, and dashed black lines indicate the MSDs of the two-species AAVM, single-species AAVM, and Brownian motion, respectively.
(c, d) Mean particle density of a single particle within the interaction radius $R$.
Color maps indicate the particle density of (c) the same-type and (d) different-type particles in the steady state. 
The coordinates were chosen such that the center particle moves in the positive direction of the abscissa.
The density is normalized by the average density of the particles within the distance $R$ (the number of type A or B particles within the interacting region divided by the area $\pi R^2$). 
The regions with higher and lower densities than the averaged density are represented by yellow- and black-color scales, respectively.
}
\label{fig:AAVM}
\end{figure}

{\it Results.---} 
Figure~\ref{fig:AAVM}(a) shows the observed time series of the two-species AAVM with the parameter $c=1.0$ (from top to bottom).
Starting from the random disordered state, the particles form several small clusters with a striped pattern. 
These clusters collide and merge with each other repeatedly, forming one large cluster in the steady state (Movie SMov.~1 in~\cite{supp}). 
The cluster moves in one direction. 
Thus, the mean squared displacement (MSD) of the particles is proportional to $\Delta t^2$ indicating ballistic behavior (Fig.~\ref{fig:AAVM}(b) green line). This contrasts with the long-term dynamics of the one-species AAVM that exhibits diffusive behavior (Fig.~\ref{fig:AAVM}(b) blue line and Supplementary Fig.~1 in~\cite{supp}).

In the cluster, each stripe consists of particles of the same type, indicating that particles of the same type aggregate and align despite their anti-aligning interactions. 
These stripes, composed of either type A or B particles, alternate in sequence parallel to the direction of cluster propulsion.
Notably, the stripe spacing $\ell$ remains nearly constant throughout the cluster and is proportional to $R$; $\ell/R$ is $0.615 \pm 0.015$, calculated through statistical analysis of $R$ incremented from $1.5$ to $3.5$ in steps of $0.5$, where the ratio $\ell/R$ is almost independent of the interaction radius. 
In addition, the width of the cluster in the steady state is approximately $2R$.
When the total number of particles is larger, instead of the width increasing, the particle density within the cluster becomes greater; alternatively, the length of the 
cluster is extended by increasing the number of stripes while maintaining a stripe spacing.
Simulation results for different total numbers of particles are presented in 
SM Sec.~S2~\cite{supp}, where a number of moving striped clusters coexist when the total number of particles is large.
In addition, even when the population ratio of type A and B particles is imbalanced, SHACO clusters still form, with the minority species accompanied by a nearly equal number of particles from the majority species (see SM Sec.~S3~\cite{supp}). The excess majority particles excluded from the clusters move randomly without directional order, similar to the behavior in single-species systems.

Reflecting the equispaced stripe patten, the average particle density within the interaction radius of a reference particle also exhibits a stripe pattern (Fig.~\ref{fig:AAVM}(c) and (d)).
Particles of the same type or different types are highly concentrated within the striped domains, which are separated by a distance $\ell$ and do not overlap.
This stripe pattern in the average density distribution remains unchanged and scales only when $R$ is changed.

These observations indicate that the inter-species interaction between the different types of anti-aligned self-propelled particles leads to a global polar order with a striped cluster.
Because such a polar state with a striped cluster has not been observed either in the one-species AAVM (see Ref.~\cite{Escaff2024} and Supplementary Fig.~1 in Ref.~\cite{supp}) or in the original Vicsek model and its variants according to the aligning interaction (see Ref.~\cite{Vicsek1995}), we propose the term SHACO~\cite{foot1} for this state.

\begin{figure}[tbhp]
\includegraphics[width=0.95\linewidth]{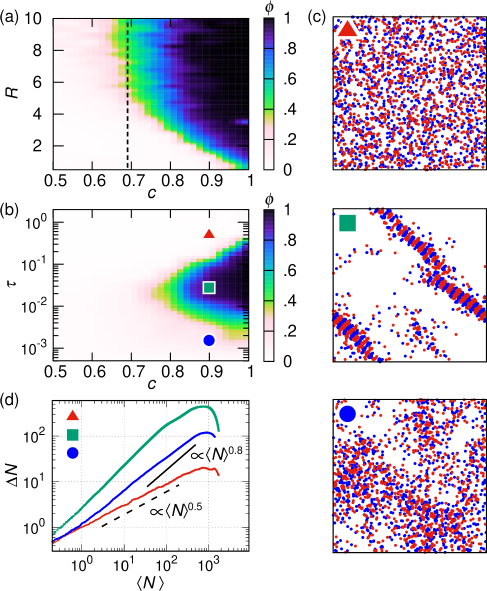}
\caption
{Polar order parameter dependence on the parameter sets of (a) $(c, R)$ and (b) $(c,\tau)$. 
Color map indicates the polar order parameter $\phi$.
The black dashed line in (a) indicates the analytically calculated transition point $c = 0.69$ (see the main text).
We used $L=64$ as the system size for calculating the order parameter.
The other parameters were fixed to the values described in the main text.
(c) Snapshots of the simulations using the parameters indicated by the symbols in (b):
$(c,\tau) = (0.9, 0.5)$ (red triangle), $(0.9, 2.81\times 10^{-2})$ (green square), and $(0.9, 1.58\times 10^{-3})$
(blue circle).
(d) Number fluctuation measured using the time-averaged particle number $\langle N \rangle$ and its standard deviation $\Delta N = \sqrt{\left(N-\langle N\rangle \right)^2}$ within a given observation region.
$\Delta N$ scales asymptotically as $\langle N \rangle^{\alpha}$.
The same simulation data as in (b) and (c) were used.
}
\label{fig:phase_diagram}
\end{figure}

In which parameter region  can this SHACO state be observed?
We measured the polar-order parameter $\phi \equiv \left| (1/N_T)\sum_{j}e^{i\theta_j}\right|$, where~$N_T=N_A+N_B$, as an indicator of the SHACO phase in various ranges of the parameter sets of $c$, $R$, and $\tau$ (Fig.~\ref{fig:phase_diagram}).
When the interaction range $R$ is small, particles leave the interaction zone before alignment can take effect. 
To compensate for this short interaction time, a stronger alignment strength  $c$ is required to establish sufficient polar order for stripe formation.
The required strength of $c$ decreases as $R$ increases, finally reaching the plateau at $c \simeq 0.7$ as indicated by the black dashed line in Fig.~\ref{fig:phase_diagram}(a).
This suggests that even if the magnitude of interactions between different types of particles does not exceed that between the same type of particles, the heterogeneous particles can facilitate the emergence of the polar order.

Even when $c$ is sufficiently large, $\tau$ must be in the specified range for the emergence of the SHACO state (Fig.~\ref{fig:phase_diagram}(b)).
In regions where $\tau$ is either too large or small, the polar order disappears in the steady state. 
Although these two states cannot be distinguished in the polar order parameter, they exhibit different behaviors in terms of the number fluctuation (Fig.~\ref{fig:phase_diagram}(c) and (d); Movie SMov.~2 in~\cite{supp}).
In the region with a small $\tau$, the particle density largely fluctuates locally and the exponent of the number fluctuation, $\alpha$, is approximately $0.8$, which is similar to the value observed in the Vicsek model and is known as the giant number fluctuation (GNF)~\cite{Toner1995,Toner1998,Chaté2008,Ikeda2024}.
A more detailed analysis including the finite-size effects is provided in SM Sec.~S4~\cite{supp}.
Note that GNF in our system is observed even though no global polar order is present. The emergence of GNF in disordered phases was reported in a previous study on active Brownian particle systems~\cite{Kuroda2023}.
In the region with a large $\tau$, the exponent is approximately $0.5$, as in the normal fluctuation.
This suggests that the phases with small and large $\tau$ are distinct, and in the small-$\tau$ phase, the characteristics of the Vicsek model are recovered even in the presence of anti-aligning interactions.
Additional data on the characteristics of the particle distribution can be found in SM Sec.~S5~\cite{supp}.
The range of $\tau$ for stabilizing the SHACO state increases as $c$ becomes larger.

\begin{figure}[tbhp]
\centering
\includegraphics[width=0.95\linewidth]{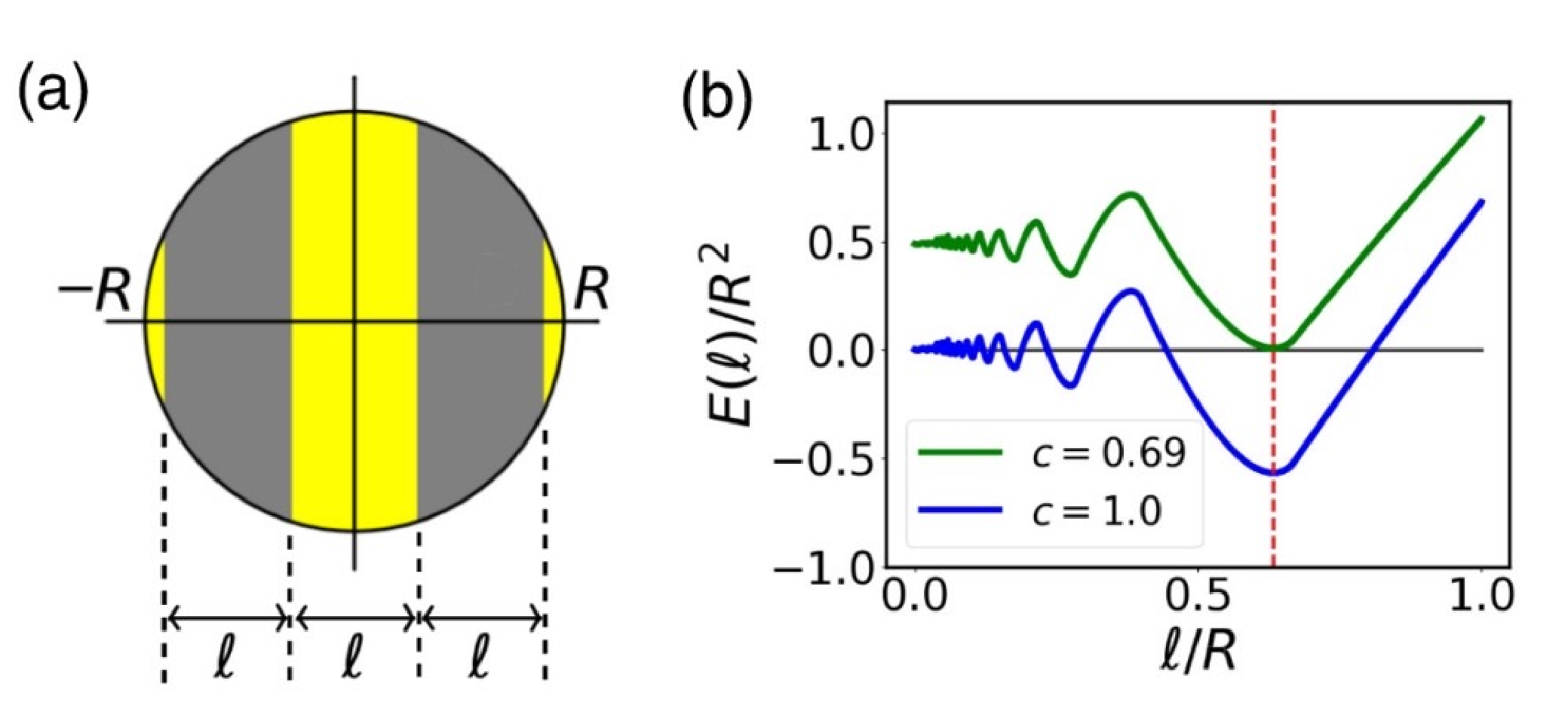}
\caption{
Mean-field analysis of the interaction energy.
(a) Distribution of particles assumed by idealizing the numerically obtained particle density (Fig.~\ref{fig:AAVM} (c) and (d)).
Yellow and gray areas represent the regions where the same-type and different-type particles exist in high concentrations, respectively.
(b) Interaction energy normalized by the area within the interaction radius, $E(\ell)/R^2$, calculated through the mean-field analysis. 
We calculated $E(\ell)/R^2$ by setting $c$ to $1.0$ (blue) and $0.69$ (green).
The red dashed line corresponds to $\ell/R = 2/\sqrt{10}$ at which $E(\ell)$ is the minimum irrespective of $c$.
}
\label{fig:mean_field}
\end{figure}

Why is the stripe pattern of the SHACO state stable, and why does the transition to the SHACO state occur even when the magnitude of the interaction between particles of different types does not exceed that between particles of the same type?
In the stable striped cluster, each stripe consists of almost only one type of particle, and the average particle distribution forms a regular striped pattern in the interaction radius of $R$, as shown in Fig.~\ref{fig:AAVM}(c) and (d).
This observation allows us to develop a mean-field theory to address the condition for the emergence of the polar order. 
We assume that the idealized particle distribution within the interaction radius of $R$ is the stripe of a different species with a constant spacing $\ell$, as illustrated in Fig.~\ref{fig:mean_field}(a).
In addition, we employ the mean-field assumption in which the particle concentration is uniform in each stripe and all particles are exposed to the same particle field.
For a given value of $\ell$, the interaction region is divided into $2n_s+1$ striped domains, where $2n_s+1$ is
the smallest odd integer greater than or equal to $2R/\ell$.
All particles are assumed to be aligned in the positive direction of the $x$-axis.
In this mean-field distribution, the direction $\theta$ of the particle at the center follows 
$\tau \dot{\theta} = E(\ell) \sin \theta$, with the interaction energy $E(\ell)$ given by 
\begin{equation}
    \label{eq:E1}
    E(\ell) = -2\int_{-R}^{R} \!\sigma_\ell(x) \sqrt{R^2-x^2}\dd{x}~,
\end{equation}
where $\sigma_\ell(x) = -1$ in the range of $(2n-1/2)\ell < x < (2n+1/2)\ell$, where $n$ is an integer, and $\sigma_\ell(x) = c$ otherwise.

The functional form of $E(\ell)$ is shown in Fig.~\ref{fig:mean_field}(b). 
$E(\ell)$ takes its minimum at $\ell = 2R/\sqrt{10} \approx 0.632 R$, which coincides with the observed stripe spacing in the clusters (see SM Sec.~S6~\cite{supp}).
This result demonstrates the smallest value of $E(\ell)$ corresponding to the most stable striped pattern was realized in the simulations.
The spacing is proportional to $R$ and is independent of the other parameters.
In addition, the minimum value of $E(\ell)$ must be negative
for the polar order to be stable.
Because the minimum value is calculated as $E(2R/\sqrt{10}) = 2R^2\left(\cos^{-1}\!\left(4/5\right)-c\sin^{-1}\!\left(4/5\right)\right)$, the transition point between the SHACO state and disordered state is evaluated as $c \approx 0.69$, which is in excellent agreement with the minimum value of the numerically obtained $c$ for the polar order to emerge (Fig.~\ref{fig:phase_diagram}(a), black dashed line).
Note that, when $R$ is small, the fluctuations in the number of particles within the interaction radius are too large to ignore.
Then, the finite-size effects cause the transition point $c$ to shift to higher values, as the reduced correlation length makes it more difficult for the particles to align.
Besides, when the particle self-propelled velocity $v_0$ is close to zero, the system behaves diffusively and does not form clusters, indicating that self-propulsion is indispensable despite the fact that the mean-field analysis is independent of $v_0$.

{\it Discussion.---}
In this Letter, we report a novel global polar-ordered state, termed the SHACO state, incorporating multiple species of anti-aligning self-propelled particles. 
The emergence of the SHACO state critically depends on the interaction strength between different types of particles, denoted by $c$, and timescale of alignment, $\tau$.
Remarkably, the SHACO state can appear even when the aligning interaction between different types of particles is weaker than the anti-aligning interaction among particles of the same type, achieved through the formation of striped clusters. 
Similar bilayer structures have been reported for passive particles with short-range attraction and long-range repulsion 
interactions~\cite{Patsahan2021,Ciach2023}, while the anti-aligning self-propelled particles exhibit cluster formation in the absence of short-range attraction.
The mean-field analysis explains the structure of the cluster and effectively captures the parameter dependence on $c$, while the dependency on $\tau$ remains an open question.
Upon examining the polar order parameter $\phi$, it appears that the SHACO state undergoes a reentrant transition with variation in $\tau$. 
However, two phases with too large or small $\tau$ can be distinguished by the emergence of a GNF (Fig.~\ref{fig:phase_diagram}(d)).
Understanding why the GNF, similar to that observed in the original Vicsek model, is observed even in the presence of anti-aligning interactions, and how the transition between the GNF and clustered phases—absent in the Vicsek model—occurs, are potential topics for future research.
In this study, we adopted positional noise, but the qualitative characteristics of the dynamics remain unaffected even when angular noise is introduced (see SM Sec.~S7~\cite{supp}).

Recent studies have demonstrated that multi-species systems of self-propelled particles give rise to a variety of complex patterns~\cite{Ouazan-Reboul2023,Dinelli2023,Duan2023,Duan2024}, including moving stripe patterns under reciprocal~\cite{Chatterjee2023} and non-reciprocal~\cite{Saha2020,Knezevic2022} interactions. However, these stripe patterns do not form clusters, and the stripes are oriented perpendicular to the direction of motion, consistent with the fact that self-propelled particles with aligning interactions typically form moving bands that extend perpendicular to their direction of motion known as Vicsek waves~\cite{Chaté2008,Gregoire2004,Yamanaka2014,Denk2020}.
In contrast, the moving cluster observed in this study elongates along the direction of motion while maintaining a constant width. This unique behavior is attributed to the conflicting interactions between particles of the same type and those of different types. 
It should be noted, however, that our present analysis does not explain why the cluster maintains a finite width.
Additional simulations starting from an initially uniform stripe state show that the particles aggregate in the lateral direction, forming elongated clusters with a finite, self-selected transverse width (see SM Sec.~S8~\cite{supp}).
Although a comprehensive analysis, including that based on theoretical approaches such as the application of Boltzmann-Ginzburg-Landau theory~\cite{Peshkov2014}, remains an open challenge, these competing
interactions provide fresh insights into the dynamics of multi-species self-propelled particles and will generate various new and complex patterns.
The emergence of such behavior within a minimal model suggests that the underlying mechanisms are robust and may inspire future investigations into more complex or biologically plausible systems.

Such conflicting interactions are expected to be observed at various levels in living systems; for example, regarding animal behaviors, individuals of the same sex repel each other, 
whereas individuals of the opposite sex attract each other, for enhancing the chances of mating.
Although it is difficult to identify the SHACO state in nature, it is much easier and more interesting to artificially design the interaction and control the state.
Recent studies succeeded in synthesizing active matter with repulsive interactions~\cite{Buttinoni2013,Kim2024,Ouazan-Reboul2023}. 
The introduction of different types of particles and the conflicting interactions between them 
are expected to lead to the emergence of the SHACO state. 
In biological systems, innovations in bacterial surface modification~\cite{Lin2023} may enable controlled formation of cell surfaces that generate such interactions.
A binary mixture system of programmable robots that analyzes neighboring individuals 
and controls their interactions (cf. Ref.~\cite{Chen2024}) is another candidate.

In conclusion, the multiple species of self-propelled particles and conflicting interactions between them
give rise to new phenomena of active matter. 
We hope that further studies, including the exploration of experimental realizations discussed above, will enrich our understanding and provide deeper insights into the underlying mechanisms of non-equilibrium systems.\\

{\it Note.---} A preprint independently a studying similar model~\cite{Lardet2025} was reported during the revision of this letter.

\begin{acknowledgments}
We are grateful to J. Yamagishi and K. Mitsumoto for their fruitful discussions. 
This study was supported by JSPS KAKENHI Grant Numbers 24H01931 (to S.I.), 21K15048 (to T.S.H.), and JST CREST Japan, Grant Number JPMJCR1923 (to S.I.).
\end{acknowledgments}

\newpage

\setcounter{figure}{0}
\setcounter{equation}{0}
\setcounter{table}{0}

\renewcommand{\figurename}{Supplementary Fig.}
\renewcommand{\tablename}{Supplementary Table}
\renewcommand{\thefigure}{S\arabic{figure}}
\renewcommand{\theequation}{S\arabic{equation}}
\renewcommand{\thetable}{S\arabic{table}}

\onecolumngrid

\begin{center}
\large{\textbf{Supplemental materials of \\`Anti-aligning self-propelled model of two species: Emergence of \textit{Self-organized Heterogeneous Aligned and Clustered Order}'}}\vspace{0.5cm}\\

\normalsize
Takahiro Oki$^{1}$, Tetsuhiro S. Hatakeyama$^{2}$, Seiya Nishikawa$^{3}$, Shuji Ishihara$^{3,4}$, Toshinori Namba$^{3,4}$\\
\begin{minipage}{0.55\linewidth}   
{\it $^{1}$Department of Physics, the University of Tokyo, \\\hspace{0.1cm} 7-3-1 Hongo, Bunkyo-ku, Tokyo 113-0033, Japan\\
$^{2}$Earth-Life Science Institute (ELSI), Institute of Science Tokyo, \\\hspace{0.1cm} 2-12-1-IE-1 Ookayama, Meguro-ku,
Tokyo 152-8550, Japan\\
$^{3}$Graduate School of Arts and Sciences, The University of Tokyo, \\\hspace{0.1cm} Komaba 3-8-1, Meguro-ku, Tokyo 153-8902, Japan \\
$^{4}$Universal Biology Institute, The University of Tokyo, \\\hspace{0.1cm} Komaba 3-8-1, Meguro-ku, Tokyo 153-8902, Japan}\vspace{0.4cm}\\
\end{minipage}
\end{center}

\section*{S1. One-species anti-aligning Viscek model}

\begin{figure*}[htbp]
\centering
\includegraphics[width=\linewidth]{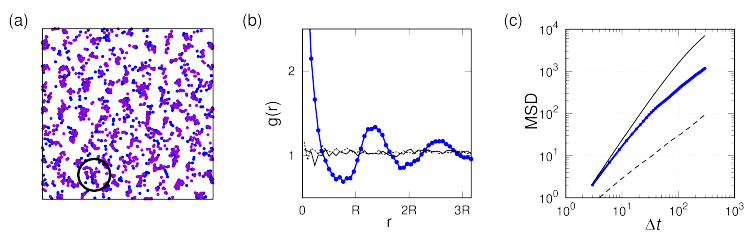}  
\caption{Characteristics of the one-species AAVM.
(a) Typical snapshot of the one-species AAVM. 
Black circles indicate interaction regions of radius, $R$, for a certain particle.
(b,c) Comparison of the one-species AAVM and the persistent random walk model (PRWM). (b) Radial distribution function, $g(r)$. (c) Mean square displacement (MSD). 
The blue lines are for the one-species AAVM. The solid and dashed black lines are for PRWM with $\eta =0.1$ and $\eta =1.0$, respectively.}
\label{fig:SFig1}
\end{figure*}

The one-species anti-aligning Vicsek model (AAVM) is given by the following equations:
\begin{align}
 \dot{\bm{r}_i} &=  \bm{s}_i + \sqrt{\sigma}\bm{\xi}_i\ , \\
    \tau\dot{\theta}_i &= \langle \sin{(\theta_i - \theta_j)} \rangle_{R}\ . \label{eq:dot_theta_i2}
\end{align}
In the one-species AAVM, particles form unstable small clusters at the steady state (Supplementary Fig.~\ref{fig:SFig1}(a) and Movie SMov3).
These clusters are separated by a characteristic distance depending on $R$, as indicated by the radial distribution function $g(r)$ that represents the density as a function of distance from a reference particle (blue line in Supplementary Fig.~\ref{fig:SFig1}(b)).
The polar order is absent.
These observations are consistent with the recent study of Ref.~\cite{Escaff2024}.
These structured particle distribution of the one-species AAVM differs from the equilibrium particle distribution of the persistent random walk model (PRWM). 
In PRWM, Eq.~\eqref{eq:dot_theta_i2} is replaced by $\dot{\theta}_i = \eta\epsilon_{i}^{t}$ where $\epsilon_{i}^{t}$ represents white noise with uniform distribution over $[-\pi,\pi]$.
Particles in the PRWM do not interact with each other, and the radial function is flat as indicated by black lines 
in Supplementary Fig.~\ref{fig:SFig1}(b) (solid line: $\eta = 0.1$, dashed line: $\eta = 1.0$).
Note that the mean square displacement of a particle in the one-species AAVM is similar to that of PRWM;
particle motion is ballistic at short timescales but becomes diffusive at longer timescales (Supplementary Fig.~\ref{fig:SFig1}(c)).

\newpage

\section*{S2. The SHACO state with larger number of particles}

We investigated the SHACO state in systems with larger numbers of particles than those studied in the main text, using the parameter set $R = 3.0$, $c = 0.9$, and $\tau=2.81\times10^{-2}$ (corresponding to the green square in Fig.~2b).
Numerical simulations were performed by changing the total number of particles $N_T = N_A + N_B$ and system size $L^2$, while keeping the particle density fixed at $\rho_0=N_T/L^2=1.0$ and maintaining an equal mixture ratio $N_A = N_B$.
Simulation snapshots for $N_T=1,600$, $10,000$ and $40,000$ are shown in Supplementary Figs.~\ref{fig:SFig2}(a-c) and are also presented in Supplementary Movie SMov4 and SMov5.
While the global order parameter $\phi$ nearly saturates by $t < 2,000$ and remains stably high thereafter (Supplementary Fig.~\ref{fig:SFig2}(d), bottom panels), multiple clusters form and coexist in large systems ($N_T = 10,000$ and $40,000$; Supplementary Fig.~\ref{fig:SFig2}(c,d) and Supplementary Movie SMov5).
These clusters continuously undergo dynamic processes such as condensation, elongation, and splitting, primarily due to collisions among them.
To quantify these dynamics, we measured spatial fluctuations in the particle number density, $\Delta\rho$, defined as the standard deviation of the local particle number computed over a regular lattice with square grid cells of size $\Delta x \times \Delta y = 4.0 \times 4.0$.
The time series of $\Delta\rho$ is shown in Supplementary Fig.~\ref{fig:SFig2}(d), revealing temporal variations that reflect repeated collisions and fragmentations of clusters, while the global alignment remains stable.
These dynamic behaviors persisted throughout the entire duration of the simulations, at least over the timescales we investigated.

\begin{figure*}[bp]
\centering
\includegraphics[width=\linewidth]{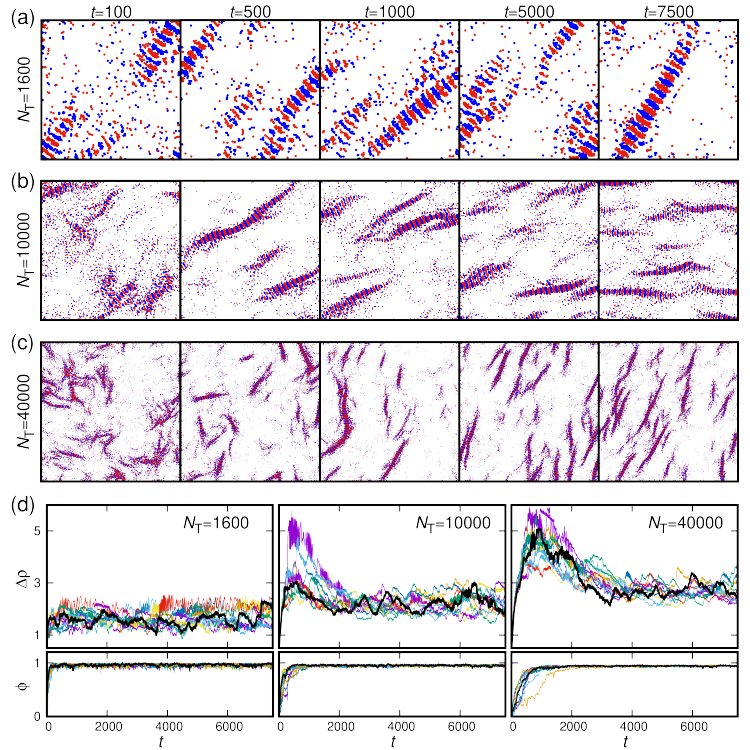}
\caption{
Cluster formation in different size of systems.
(a–c) 
Simulation snapshots at $t = 100$, $500$, $1,000$, $5,000$, $7,500$, for different total particle numbers. (a) $N_T=1,600$, $L=40$, (b) $N_T=10,000$, $L=100$, (c) $N_T=40,000$, $L=200$.
The particle density is fixed at $N_T/L^2=1.0$, with parameters $R=3.0$, $c=0.9,$ and $\tau = 2.81\times 10^{-2}$. 
(d) Time series of the local density fluctuation $\Delta\rho$ for $N_T = 1,600$, $10,000$, $40,000$.
Data from $11$ independent runs are shown for each total particle number.
The trajectories corresponding to the simulations in panels (a–c) are highlighted as thick black lines.
The bottom panels show the time series of global polar order parameter, $\phi(t)$.
}
\label{fig:SFig2}
\end{figure*}

\clearpage

\section*{S3. Unequal particle number ratios between species}
We performed simulations in which the fraction of type A particles (blue particles) was varied, with the total number of particles fixed at~$N_T=N_A+N_B=2,000$.
Snapshots from these simulations are shown in supplementary Fig.~\ref{fig:SFig3}(a) and Supplemental Movie SMov6.
The particles of minor species form SHACO cluster together with a nearly equal number of particles from the majority species, showing ballistic migration. 
The remaining particles of the majority species, which are excluded from SHACO, move randomly without exhibiting directional order, similar to the behavior observed in single-species simulations.
The polar-order parameter $\phi = \left| (1/N_T)\sum _je^{i\theta_j}\right|$ for various values of $N_A$ is shown in Supplementary Fig.~\ref{fig:SFig3}(b), in which
$\phi$ exhibits the highest value at even mixture $N_A = N_B$.
However, when focusing on the polar-order parameter of A-type participles, $\phi_A =  \left| (1/N_A)\sum _{j}e^{i\theta_j}\right|$ where sum is taken over A-type particles,  
$\phi_A$ takes almost unity between finite range of $N_A (< N/2)$ (Supplementary Fig.~\ref{fig:SFig3}(c)).
Thus, in the uneven mixture, minority species is responsible for formation of the SHACO cluster and exhibit high order parameters.

\begin{figure*}[htbp]
\centering
\includegraphics[width=\linewidth]{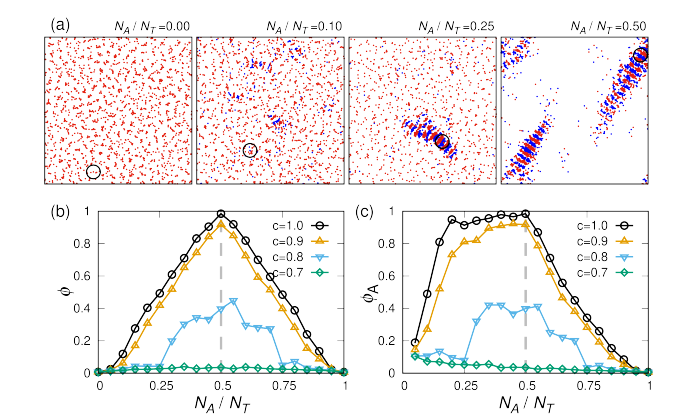}  
\caption{
Alignment orders in AAVM with unequal type-A/type-B ratio $N_A/N_B$.
(a) Snapshots at time $t = 750.0$ for $N_A/N_T = 0.00, 0.10, 0.25$, and $0.50$.
The black circle indicates the interaction region of radius $R$.
(b) Global order parameter, $\phi$. 
(c) Order parameter for 
particle of type A, $\phi_A$. 
Results with different $c$ are shown.
}
\label{fig:SFig3}
\end{figure*}

\clearpage

\section*{S4. Finite-size analysis of GNF}

We investigate the finite-size effect for GNF observed in Fig.~2(d) in the main text.
Using the parameter set indicated by the blue circle in Fig.~2(b) ($R = 3.0$, $c = 0.9$, and $\tau = 1.58 \times 10^{-3}$), we performed numerical simulations with varying system sizes while keeping the particle density fixed ($N_T / L^2 = 1.0$).
The snapshot at $N = 40,000$ is shown in Supplementary Fig.~\ref{fig:SFig4}(b) (see also Supplemental Movie SMov7).
The number fluctuation is presented in Supplementary Fig.~\ref{fig:SFig4}(a).
Regardless of the total number of particles, the scaling exponent of the particle number fluctuation consistently surpass $0.8$,
where the region exhibiting the exponent is more prominent as the total number of particles increases.
The fitted exponent is approximately $0.839$, which is close to $0.84$ derived in the ordered phase of the standard Vicsek model~\cite{Ikeda2024}.
Note that GNF in our system is observed even though no global polar order is present. 
The emergence of GNF in disordered phases was reported by previous study on active Brownian particle systems~\cite{Kuroda2023}.

\begin{figure*}[htbp]
\centering
\includegraphics[width=\linewidth]{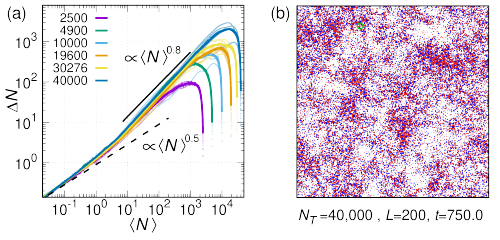}  
\caption{
Verification of GNF by the finite-size analysis.
(a) Number fluctuation measured using the time-averaged particle number $\langle N \rangle$ and its standard deviation $\Delta N = \sqrt{\left(N-\langle N\rangle \right)^2}$ within a given observation region, at the parameter of blue circle in Fig.~2(b,d) in the main text 
($R=3.0, c=0.9$, and $\tau=1.58\times10^{-3}$).
The result with different total number of particles, 
$N_T =~2,500,~4,900,~10,000,~19,600,~30,276$, and~$40,000$.
are shown. For each $N_T$, results from four independent simulation runs are plotted by small dots and their average are shown by thick line.
(b) Snapshot of the simulations 
at $N_T = 40,000$. 
The green circle indicates the interaction region of radius $R$.
}
\label{fig:SFig4}
\end{figure*}

\clearpage

\section*{S5.Characterization of particle distribution in ordered and disordered states}

\begin{figure*}[htbp]
\centering
\includegraphics[width=0.905\linewidth]{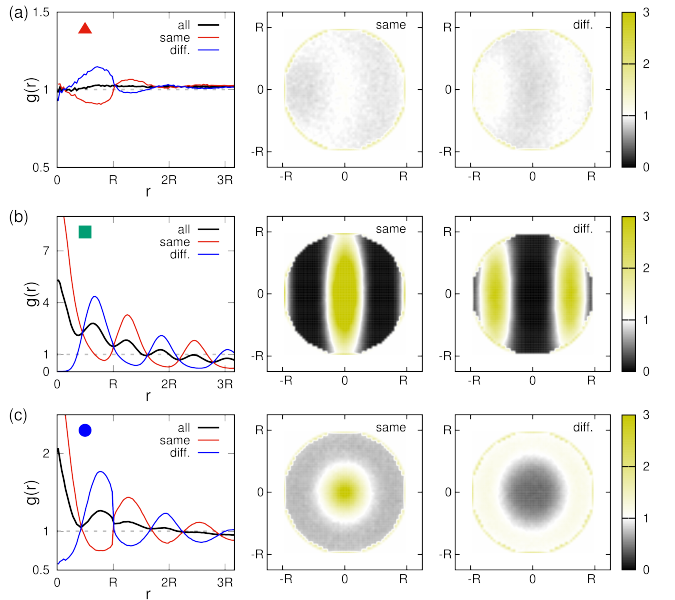}  
\caption{
Radial distribution function $g(r)$ and the mean particle densities around a single particle at the parameter sets represented by symbols in Fig.~2(b–c).
(a) $\tau=0.5$ (red triangle), (b) $\tau=2.81\times 10^{-2}$ (green square), and (c) $\tau=1.58\times 10^{-3}$ (blue circle).
For each panel (a-–c), the left figure shows the radial distribution function $g(r)$.
The center and right figures show the mean particle densities around a single particle within the interaction radius $R$ for same-type and different-type particles, respectively.
}
\label{fig:SFig5}
\end{figure*}

In the parameter region showing no polar order, GNF was observed in small $\tau$ region whereas was not in large $\tau$ region (Fig.~2(b) in the main text). 
To further investigate the difference between them, we characterize their spatial structures by remeasuring the radial distribution function (RDF) and the mean particle density around a single particle within the interaction radius~$R$.
Supplementary Fig.~\ref{fig:SFig5} shows the RDF for the three parameter sets illustrated in Fig.~2(c) of the main text.
The RDF for $\tau=0.5$ (red triangle in Fig.~2(b) in the main text) is almost flat, with a slight excess of different-type pairs within the interaction range~$R$ (Supplementary Fig.~\ref{fig:SFig5}(a)). 
In contrast, the RDF for~$\tau=1.58\times 10^{-3}$ (blue circle) exhibits signature of spatial pattern (Supplementary Fig.~\ref{fig:SFig5}(c)).
This indicates that, although the overall arrangement of particles fluctuates due to the lack of global polar order, the local density forms pattern around each particle (see Supplementary Fig.~\ref{fig:SFig4}(b)).
The numerical simulation indicates that the stripe pattern of type A and B particles appear locally and transiently (see Supplemental Movie SMov7).
Under the SHACO-forming conditions, different-type particles are rarely found in close proximity (Supplementary Fig.~\ref{fig:SFig5}(b)). 
In contrast, under low-$\tau$ conditions where GNF is observed, different-type particles are observed to coexist within the same local neighborhood (Supplementary Fig.~\ref{fig:SFig5}(c)).

\clearpage

\section*{S6. Analysis of Interaction Energy, $\mathbf{E(\ell)}$}

This section details the analysis of Eq.~(3) in the main text. 
In the mean-field analysis, we assumed that particles around a reference particle
are distributed in stripes along the axis of moving direction within the interaction region of radius $R$ (Fig.~3(a)). 
Each stripe domain consists of either type A or type B particles, with a uniform density within each domain.
Let $\ell$ denote the width of the stripe, which serves as the control parameter in our analysis.
For a given $\ell$, the interaction region contains $2n_s+1$ stripes of width $\ell$ along with two additional stripes of shorter width on both sides. Here, $n_s$ is provided by
\begin{align}
    \label{eq:nsl}
    n_s = \lfloor R/\ell - 1/2 \rfloor\ ,
\end{align}
with the floor function $\lfloor x \rfloor$, the greatest integer less than or equal to~$x$.
The center stripe containing the origin consists of the same type particles as the reference at the origin.
Then Eq.~(3) in the main text is written as
\begin{align*}
    E(\ell)
    = 
    \sigma_{-n_s-1}  \int_{-R}^{-(n_s+\frac{1}{2})\ell} \sqrt{R^2-x^2} \dd{x}
    +\sum_{n=-n_s}^{n_s} \sigma_{n}
    \int_{(n-\frac{1}{2})\ell}^{(n+\frac{1}{2})\ell}
    \sqrt{R^2-x^2} \dd{x} + 
    \sigma_{n_s+1}\int_{(n_s+\frac{1}{2})\ell}^R
    \sqrt{R^2-x^2} \dd{x}\         ,
\end{align*}
where~$\sigma_\ell(x)$ is replaced by $\sigma_n=1-c + (-1)^{n}(1+c)$. By using the primitive function of $\sqrt{R^2-x^2}$,
\[ G(x) \equiv \int_{0}^x \sqrt{R^2-t^2}dt = \frac{R^2}{2} \left(\frac{x}{R}\sqrt{1-\left(\frac{x}{R}\right)^2} + \sin^{-1}\left(\frac{x}{R}\right) \right)~, \]
and referring to the relationship $\sigma_{n+1}-\sigma_{n}=2(-1)^{n+1} (1+c)$, $\sigma_{n_s+1}+\sigma_{-n_s-1}=2\sigma_{n_s+1}$, and 
$G(x) = -G(-x)$, $E(\ell)$ is simplified as follows:
\begin{align}
    E(\ell)
    &= - \sigma_{-n_s-1}G(-R)        
    -\sum_{n=-n_s-1}^{n_s} 
    \left(\sigma_{n+1}-\sigma_{n}\right)G\left(\left(n+\frac{1}{2}\right)\ell\right) 
    +\sigma_{n_s+1}G(R)\ , \nonumber\\    
    &=2(1+c)\sum_{n=-n_s-1}^{n_s} 
    (-1)^n G\left(\left(n+\frac{1}{2}\right)\ell \right)+2\sigma_{n_s+1}G(R)\ ,\nonumber\\    
    & = 4(1+c)\sum_{n=0}^{n_s}  (-1)^n G\left(\left(2n+1\right)\frac{\ell}{2} \right)+\frac{\pi}{2}R^2 \sigma_{n_s+1}~.    \label{eq:4}
\end{align}
Here we used $G(R)= \pi R^2/4$.
As shown in Fig.~3(b), $E(\ell)$ exhibits oscillatory damping towards zero as the width~$\ell$ of the stripe region decreases.

Next, we calculated the minimum value of $E(\ell)$ at which the cluster is most stable
(more precisely, stable to the angular perturbation to the reference particle).
Using $G'(x) = \sqrt{R^2-x^2}$, derivative of $E(\ell)$ with respect to $\ell$ yields
\begin{align} 
    \label{eq:6}
    \dv{E(\ell)}{\ell} = 2(1+c)R\sum_{n=0}^{n_s} (-1)^n \left(2n+1\right) \sqrt{1 - \left(2n+1\right)^2\left(\frac{\ell}{2R}\right)^2}\ . 
\end{align}
As evident from Eq.~\eqref{eq:6}, the solution $\ell$ to $dE(\ell)/d\ell=0$ is independent of $c$, and consequently, $\ell$ where $E(\ell)$ reaches its minimum is also independent of $c$.

Our observation suggests that the minimum of $E(\ell)$ is realized at $n_s=1$, for which Eq.~\eqref{eq:6} is reduced to the equation
\begin{align*} 
    \sqrt{1 - \left(\frac{\ell}{2R}\right)^2} - 3\sqrt{1 - 3^{2}\left(\frac{\ell}{2R}\right)^2} =0~.
\end{align*}
The solution 
\[ \ell=\frac{2R}{\sqrt{10}} \simeq 0.63R\]
is compatible with the prerequisite $n_s=1$ (see Eq.~\eqref{eq:nsl}).
As mentioned in the main text, this width well agrees with the experimental observation.
Substitution of $\ell = 2R/\sqrt{10}$ and $n_s=1$ into $E(\ell)$ in Eq.~\eqref{eq:4} leads to
\begin{align*}    
    E(2R/\sqrt{10}) &= 4(1+c)\left[ G(R/\sqrt{10}) - G(3R/\sqrt{10})\right] + \pi R^2 \\
        & = -2(1+c)R^2 \sin^{-1}(4/5) + \pi R^2\\
        &=2R^2 \left( \cos^{-1}\left( 4/5\right) - c \sin^{-1}\left( 4/5\right)\right)~,
\end{align*}
where we used the formulae $\sin^{-1}(\alpha) - \sin^{-1}(\beta) = \sin^{-1}\left( \alpha \sqrt{1-\beta^2}-\beta\sqrt{1-\alpha^2}\right)$ and $\pi/2 = \sin^{-1}(\alpha) + \cos^{-1}(\alpha)$.
In the main text, we assessed the stability condition $E(2R/\sqrt{10})<0$ by changing $c$
and found the critical value $c \approx 0.69$.

\section*{S7. Noise Effect on the angle evolution equation}

In the main text we studied systems where noise is added to the positional variable ${\bm r}_i$ (Eq.~1 in the main text). 
Here we study the case where noise is added to the angular variable $\theta_i$, given as
\begin{align}
\dot{\bm{r}_i} &= v_0\bm{s}_i\ , \label{eq:dot_ri_wo_noise}\\
\tau \dot{\theta}_i &= - \langle m_{ij}\sin{(\theta_i - \theta_j)} \rangle_{R}+ \eta\ U(-\pi,\pi) \ ,\label{eq:dot_theta_wt_noise}
\end{align}
where $U(-\pi,\pi)$ represents a random variables from a uniform distribution with support~$(-\pi,\pi)$, and $\eta$ indicates the strength of the noise.
The polar-order parameter $\phi$ is plotted  against the strength of noise in Supplementary Fig.~\ref{fig:SFig6}(a). 
The polar-order is maintained as long as the noise intensity is small, as expected.
The snapshot at $\eta=0.04$ is shown in Supplementary Fig.~\ref{fig:SFig6}(b).

\begin{figure*}[htbp]
\centering
\includegraphics[width=\linewidth]{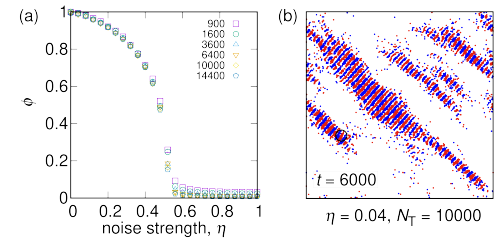}  
\caption{Effect of noise on the angle evolution equation.
(a) Global order parameter as a function of noise strength~$\eta$ for total particle numbers $N_T=900,1,600,3,600,6,400,10,000$, and $14400$.
The particle density is fixed at $N_T/L^2=1.0$, with parameters $R=3.0, c=1.0,$ and~$\tau=0.1$.
(b) Snapshot at $t=6,000$ in the case of $\eta=0.04$ and~$N_T=10,000$. 
}
\label{fig:SFig6}
\end{figure*}

\clearpage

\section*{S8. Lateral instability of system-wide stripe pattern\\ and formation of SHACO clusters}

\begin{figure*}[!b]
\centering
\includegraphics[width=0.95\linewidth]{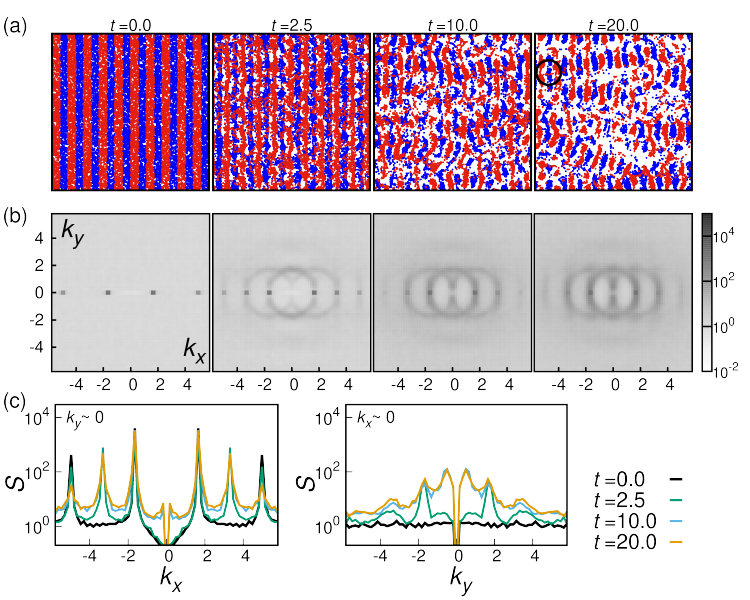} 
\caption{Temporal changes in the structure factor starting from a stripe-patterned initial state.
(a) Typical simulation snapshots at $t = 0$, $2.5$, $10$, $20$, and the number of particles are $N_A=10,000$ and $N_T=20,000$.
(b) Two-dimensional maps of the structure factor $S(\bm{k},t)$ for type A particles at time points corresponding to the snapshots in (a).
The resulting values are displayed as a grayscale, with the horizontal and vertical axes corresponding to the wave vectors $k_x$ and $k_y$, respectively.
(c) Line profiles of the structure factor corresponding to the two-dimensional maps in (b) show cross sections along $k_x\sim 0$ and $k_y\sim 0$, respectively.
}
\label{fig:SFig7}
\end{figure*}

To gain insight into how the SHACO cluster with a finite width develops,
we numerically investigated the temporal evolution of the system starting from an initially stripe-patterned particle distribution (Supplementary Fig.~\ref{fig:SFig7}(a)).
In the simulation, the system size was set to $L=20\ell_R$ along each axis.
The initial particle positions were uniformly distributed along the $y$-axis,
whereas along the $x$-axis, type-A and type-B particles were alternately arranged in stripes of width $\ell_R = 2R/\sqrt{10}$.
The particle orientations $\theta_i$ were initialized to values close to zero,
with small random perturbations drawn uniformly from the interval $(- \pi/200, \pi/200)$. 
The parameters were set to $R = 3.0$, $c = 0.9$, and $\tau=2.81\times10^{-2}$, corresponding to the green square in Fig.~2b.

As shown in the snapshots in Supplementary Fig.~\ref{fig:SFig7}(a), the initially stripe-patterned particle distribution evolves into SHACO clusters.
Notably, even at $t=20$, when SHACO clusters with a typical width of approximately $2R$ have formed, the number of stripes remains about $20$, the same as in the initial state. This observation indicates that the particles primarily undergo transverse aggregation.
To quantify the temporal evolution of the particle configuration, we tracked the structure factor of particle density at each time point $t$.
Supplementary Figs.~\ref{fig:SFig7}(b) shows the structure factor $S(\mathbf{k}, t)$ calculated from the positions of type-A particles.
The line profiles along the $k_x$ and $k_y$ wavenumbers axes 
are shown in Supplementary Fig.~\ref{fig:SFig7}(c).
Along the $k_x$ direction, multiple sharp peaks in the initial state are maintained, reflecting stability of the periodic stripe arrangement along the $x$-axis (it should be noted that at $t = 0$, the even-order harmonic modes are absent owing to the specific nature of the initial condition).
In contrast, along the $k_y$ direction, several peaks arise right after the initial state, indicating transverse instability with characteristic wavenumber of $k^\ast_y = 1.66$.
The corresponding characteristic length $2\pi/k^\ast_y = 3.79 $ is close to {twice stripe width}.
These results demonstrate that particles starting from a uniform stripe configuration undergo a transverse instability, which leads to the formation of clusters with a finite width and results in longitudinal elongation only in the SHACO state.

\clearpage

\section{SUPPLEMENTAL MOVIES}
\noindent\textbf{SMov1}\ :\  The behavior of the two-species AAVM corresponding to Fig.~1.
(a) Time evolution of the motion of the particles. 
(b, c) Mean particle densities around a single particle within the interaction radius~$R$.
Color maps indicate the particle density of (b) the same-type and (c) different-type particles.\\

\noindent\textbf{SMov2}\ :\  
Simulations of the two-species AAVM correspinding to Fig.~2 (c).
Parameters are set as $(c,\tau) = (0.9, 0.5)$ (red triangle), $(0.9, 2.81 \times 10^{-2})$ (green square), and $(0.9, 1.58 \times 10^{-3})$ (blue circle). \\

\noindent\textbf{SMov3}\ :\  The behavior of the one-species AAVM corresponding to Supplementary Fig.~\ref{fig:SFig1}(a).
(a) Time evolution of the notion of the particle.
(b) Mean particle density around a single particle within the interaction radius~$R$.\\

\noindent\textbf{SMov4}\ :\  
Simulations of the two-species AAVM shown in Supplementary Fig.~\ref{fig:SFig2}(b).
The parameter sets $(N_T, c, \tau) = (1.0\times 10^{4}, 0.9, 2.81 \times 10^{-2})$, corresponds to the green square in Fig. 2(b) of the main text.\\

\noindent\textbf{SMov5}\ :\  
Simulations of the two-species AAVM shown in Supplementary Fig.~\ref{fig:SFig2}(c).
The parameter sets $(N_T, c, \tau) = (4.0\times 10^{4}, 0.9, 2.81 \times 10^{-2})$, corresponds to the green square in Fig. 2(b) of the main text.\\

\noindent\textbf{SMov6}\ :\  Simulations of the two-species AAVM for different particle number ratios shown in Supplementary Fig.~\ref{fig:SFig3}.
Evolution in time of the motion of the particles in the ratio of $N_A/(N_A+N_B)=0.0,0.1,0.25,0.5$.\\

\noindent\textbf{SMov7}\ :\  
Simulations of the two-species AAVM shown in Supplementary Fig.~\ref{fig:SFig4}(b).
The parameter set $(N_T, c, \tau) = (4.0\times 10^{4},0.9, 1.58 \times 10^{-3})$, corresponds to the blue circle in Fig. 2(b) of the main text.\\

\clearpage

\end{document}